# Energy Efficient Neural Network Embedding in IoT over Passive Optical Networks

**Mohammed Moawad Alenazi, Barzan A. Yosuf, Taisir El-Gorashi, and Jaafar M.H.Elmirghani**
*School of Electrical Engineering, University of Leeds, LS9JT, United Kingdom*

**ABSTRACT**

In the near future, IoT based application services are anticipated to collect massive amounts of data on which complex and diverse tasks are expected to be performed. Machine learning algorithms such as Artificial Neural Networks (ANN) are increasingly used in smart environments to predict the output for a given problem based on a set of tuning parameters as the input. To this end, we present an energy efficient neural network (EE-NN) service embedding framework for IoT based smart homes. The developed framework considers the idea of Service Oriented Architecture (SOA) to provide service abstraction for multiple complex modules of a NN which can be used by a higher application layer. We utilize Mixed Integer Linear Programming (MILP) to formulate the embedding problem to minimize the total power consumption of networking and processing simultaneously. The results of the MILP model show that our optimized NN can save up to 86% by embedding processing modules in IoT devices and up to 72% in fog nodes due to the limited capacity of IoT devices.

## 1. INTRODUCTION

The uptake of the Internet of Things (IoT) is increasing at unprecedented levels across a wide variety of domains in our daily lives, primarily due to the manufacturing advancement with respect to a reduction in cost, size, and power consumption of next-generation low-power radio transceivers and microcontrollers [1]. The number of IoT devices is estimated to be between 26 billion to 50 billion devices [2]. The sheer number of IoT devices leads to the generation of massive amounts of data which is usually transported over multiple domains of the network towards the centralized cloud data center for processing to extract knowledge from the data [3] – [6]. The costly overhead of the transport network created a need for processing the collected data closer to the IoT end-devices, hence fog computing can fill this void by complementing the cloud and extending its services to the edge of the network and even further into the IoT devices [7] – [9]. Researchers have developed a chip that increases processing speed of computations in neural networks with impact on reducing the energy consumed between 94% to 95% which supports their use in smartphones and also in home appliances [10]. In the past, most of the IoT applications aimed at passive data collection and monitoring, however, recently, actuation has received lots of attention. Through the coupling of sensors and actuators that are capable of interacting with the physical world, it becomes possible for next-generation IoT based systems to perform sophisticated tasks in an automated and dynamic manner [11]. One of the applications as an example is an IoT based smart home in which sensors and actuators are coordinated intelligently to control given parameters such as energy usage based on the time of the day [12]. The topology of a neural network is based on three layers: 1) input, 2) hidden, and 3) output. As shown in Figure 1, the input nodes are all connected to each hidden node via communication links, and the data generated by the hidden nodes are fed into the output node(s) for actuation based on the weight of edges and bias values of hidden nodes [13]. The nodes within a NN are called neurons, which require processing to transform the input measurements into a desired output for the actuating nodes. Usually the edges between the neurons are needed to establish communication and synchronization for NN requests [14]. Conventionally, the data processing is mostly offloaded to a centralized cloud data center in which the input nodes' data is routed through the local gateway towards remote servers located deeper into the network and once knowledge is extracted from the processed data, the required output signal is returned back to the IoT local gateway and then this signal is used by the actuator devices. Recently, researchers have paid much attention to ANN embedding in IoT based networks such as WSNs. The authors of [15] considered the use of the processing capabilities of low-power, cheap IoT nodes and their linkages, where the IoT nodes maybe in a mesh topology, to implement an ANN. This approach realizes the fact that IoT nodes may be constrained by computational capacity and therefore may not be able to implement a full (deep) neural network at each IoT node. The work was extended in [12] by enriching their optimization framework to consider low-power routing protocols, accounting for the energy consumption of the communication in the ANN in order to optimize the allocation of hidded neurons and consequenctly improving the network lifetime. Their solution was also simulated on an IoT testbed to validate the feasibility of their concept together with measuring its performace against the centralized gateway in terms of latency. In this work, we build on the concepts in [12], [15] and extend the work by i) considering multiple IoT fields, hence data processing on aggregation nodes such as gateway fog and access fog tackles the embedding problem from another angle; (ii) introducing a practical network architecture that links the IoT fields, composed of PON elements, gateways and relays; (iii) introducing a focus on energy efficiency captured as the objective of our MILP formulation; (iv) casting the problem into a service oriented architecture, hence allowing these ANN embeddings to be included in a framework that can enable the embedding of other services such as security services (where the nodes may be interlinked motion sensors, processors and actuators such as alarms), energy saving services (where the nodes may include interlinked motion sensors, processors and actuators such as networked switches and displays.

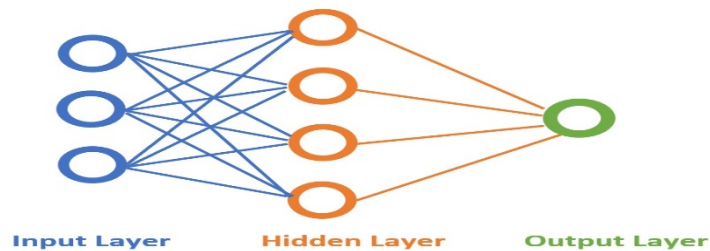

Figure 1. Neural Network Topology

The work in this paper benefits from our previous proposals for improving energy efficiency in service embedding in IoT and core networks [16] – [22], server disaggregation in data centers [23], content distribution and big data processing [23] - [28], and core, edge and fog processing in networks [29] – [34]. The remainder of this paper is organized as follows: In Section 2, we present the EE-NN embedding framework over the PON access network, Section 3 discusses the performance evaluation and the results. Finally, Section 4 provides the conclusions for the paper.

## 2. ENERGY EFFICIENT NN EMBEDDING FRAMEWORK OVER PON ACCESS NETWORK

In IoT based smart homes, different IoT devices are needed to monitor several variables inside a home such as door lights, fire and smoke detectors, and temperature sensors [35]. To enhance the smart home applications, the control system should make decisions and proactively execute different tasks through the predictive features of neural networks (NNs). This paper aims to propose a framework for intelligent IoT based smart homes by i) embedding NN service requests into the physical infrastructure of an end-to-end IoT infrastructure supported by the concept of the fog, ii) evaluating several variations of processing platforms within the IoT, Fog, and cloud in terms of the total power consumption, iii) imposing processing placement to take place in predefined locations and lifting this limitation by allowing the MILP model to choose the optimal solution for the given scenarios. The NN service requests follow the idea of the Service Oriented Architecture (SOA) and such requests are made of a virtual topology that consists of virtual nodes and links [36], [37]. The virtual links represent the required communication between the virtual nodes of a NN request within the IoT network(s) and each virtual node requires processing. We consider that a NN request involves virtual nodes spanning over two different IoT networks hence the set of links to be optimized is only feasible in the IoT architecture due to its mesh topology. The evaluated framework as shown in Figure 2 is composed of three main layers; 1) Physical layer, 2) Network layer, and 3) the Application layer. The physical layer is composed of generic low-power IoT nodes that are connected via the Zigbee protocol [38]. Multiple IoT networks are connected through their relay devices to their respective access points (APs) mounted onto the ONU devices. The application layer in our model is considered as a set of virtual nodes within a NN service request while the network layer consists of the actual transport network which includes the PON access network elements (ONU and OLT) and a single metro network ethernet switch. There is also networking infrastructure inside processing nodes such as the Access Fog and the Cloud due to the location and size of these facilities.

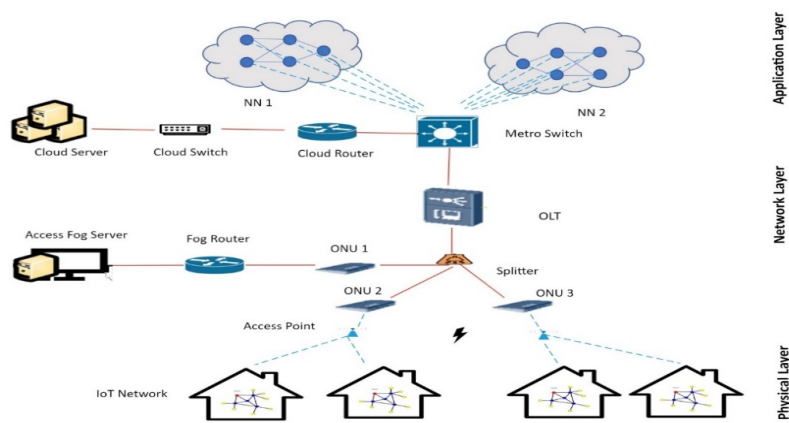

Figure 2. The evaluated architecture for the NN service embedding over the Passive Optical Network (PON).

Moreover, the framework considers the following assumptions:
- Virtual nodes within a NN request span multiple IoT networks, hence the PON access network is always required to be active in order to achieve the communication between connected virtual nodes. A virtual node's service request is represented by typical capacities of an IoT node for both processing and networking and this ranged from low to high workloads. Although the processing workload is varied, we have assumed the traffic data rate to remain constant as the same task may require different processing intensities.
- The framework considers multiple functions such as sensing functions, which are smart thermostats, and motion sensors. In addition, it has one control function and actuation functions which include alarms, actuated blinds.
- Each IoT node is mapped with one function of each type and each virtual node requests one function only.
- We have assumed that our framework accounts for a portion of the total idle power of those networking devices that have the potential to be shared by many users and applications such as ONUs, OLTs, Metro switches, Cloud LANs, etc.

The results were obtained using AMPL/CPLEX software running on a high-performance computer with 16 cores processor and 256 GB of RAM. All of the parameters used in the MILP model, both networking and processing, are summarized in Table 1 and Table 2.

| Device Type | Idle Power (W) | $\delta$ (%) | Max Power (W) | Location |
|---|---|---|---|---|
| IoT (RPi Zero) [39] | 0.5 [39] | - | 3.96 [39] | IoT |
| IoT (CC3100MOD) [40] | 0.001 [16] | - | 0.11 [16] | IoT |
| Wi-Fi Access Point | 0.34 [39] | - | 0.56 [39] | Network |
| ONU | 9 [39] | 1, 5 or 10 | 15 [39] | Gateway Fog |
| OLT | 60 [39] | 1, 5 or 10 | 1940 [39] | Network |
| Fog Router | 11.7 [39] | 1, 5 or 10 | 30 [39] | Access Fog |
| Metro Switch | 128 [39] | 1, 5 or 10 | 247 [39] | Network |
| Cloud Router | 27 [39] | 1, 5 or 10 | 30 [39] | Cloud |
| Cloud Switch | 128 [39] | 1, 5 or 10 | 423 [39] | Cloud |

Table1: Network devices capacity and power consumption parameters used for the MILP.

| Device Type | Capacity (MIPS) | W/MIPS | Idle Power (W) | Max Power (W) | Location |
|---|---|---|---|---|---|
| IoT (RPi Zero)[15] | 1000 [39] | $3460\mu$ [39] | 0.5 [39] | 3.96 [39] | IoT |
| IoT (CC3100MOD)[26] | 856 [16] | 0.856 [16] | 0.001 [16] | 0.11 [16] | IoT |
| Gateway Fog | 2400 [39] | $4375\mu$ [39] | 2 [39] | 12.5 [39] | Gateway Fog |
| Access Fog Server | 34200 [39] | $1111\mu$ [39] | 57 [39] | 95 [39] | Access Fog |
| Cloud Server | 108000 [39] | $481\mu$ [39] | 78 [39] | 130 [39] | Cloud |

Table 2: Processing devices capacity and power consumption parameters used for the MILP

## 3. RESULTS AND EVALUATIONS

In our topology we considered two separated IoT networks that are connected to the cloud via a PON access network. Each IoT network consisted of 30 IoT devices and 2 relays. In addition, the IoT networks are each connected to a single ONU and a single OLT aggregates traffic from both ONUs. The distribution of the IoT devices in both networks is random and uniform. All devices in the IoT network communicate via the Zigbee protocol (IEEE 802.15) while the relay devices are connected to the ONUs via the Wi-Fi protocol. On the other hand, the ONUs are connected to the OLT device through an optical fiber link. Our model accounts for the total power consumption which consists of two parts: 1) network power consumption and 2) processing power consumption. We consider both uplink and downlink traffic as both types of communication are needed to achieve a neural network spanning multiple IoT networks. Consequently, we allow traffic to pass from one network to another through the OLT device. It has been shown from our previous work [16], [39], [41] that the fog approach which allows for hosting processing services closer to the IoT-end devices saves a considerable amount of power compared to the traditional centralized cloud approach. To this end, our fog approach considered 3 layers of processing namely the IoT, Gateway Fog, Access Fog layers as well as the cloud. We assumed that each NN is composed of five virtual nodes that are connected through three layers that involve input nodes, hidden nodes, and the output node(s). Each node within a NN request has a task requirement consisting of processing in MIPS and traffic in kbps. All requests are assumed to have 100% SLA since task blocking is not considered, however, nodes' processing and networking capacity must be respected at all times. We have considered a linear power profile to calculate both networking and processing power consumptions and it consists of two parts: a) idle power and b) load proportional power. In order to undertake fair evaluations, we have assumed that the IoT application in this work is only responsible for a portion of the idle power ($\delta$) of the high-capacity networking devices since such devices can be shared between many users and applications [39]. Thus, we considered three scenarios in which the value of $\delta$ was changed to 1% in scenario #1, 5% in scenario #2, and 10% in scenario #3. In each scenario, a range of homogenous processing demands between 20% - 100% of the IoT device's processing capacity (MIPS) is considered. We evaluated different variations of processing in the proposed framework; 1) processing to take place only in the IoT layers, 2) processing to take place in the IoT and PON, 3) processing to take place only in PON, 4) allowing MILP model to choose processing location (optimal solution) and finally 5) all processing to take place in the centralized cloud (baseline). It is worth noting that variation 1-3, processing tasks are forced to be distributed among the relevant layers to account for circumstances where not all fog nodes have the right software package to process all the tasks. Also, evaluating different values of $\delta$ is valid as this can represent the growth of the current application. Figure 3 shows the total power consumption of scenario #1 – to scenario #3 while Figure 4 shows the total power savings (in %) of all the processing placement variations compared to the cloud. The IoT approach produces the highest savings of up to 86% across all scenarios due to the negligible power consumption of the low-power microcontroller type devices in the IoT layer. The IoT + PON approach achieved savings of up to 68% in scenario #1 due to the low value of $\delta = 1\%$ whilst this increased up to 80% in scenario #3 due to the expensive network overhead to get to the cloud, hence $\delta = 10\%$. On the other hand, the PON approach produced better savings than the IoT + PON due to better utilization of the PON fog servers and these savings were up to 72% and 84% for scenario #1 and scenario #2, respectively. In the last approach (MILP), we allowed the model to choose the optimal location for processing in all scenarios and as can be seen in Figure 4, the optimal allocation is the IoT approach.

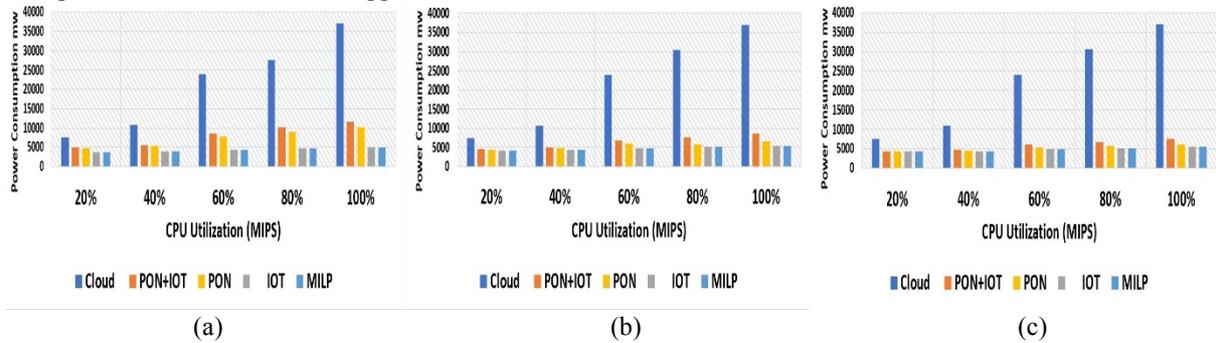

| (a) | (b) | (c) |

Figure 3. Total power consumption of the fog approach at different values of $\delta$ which represents the proportion of the idle power consumption attributed to the IoT application for high-capacity networking devices: a) when the IoT $\delta = 1\%$, b) when $\delta = 5\%$, and c) when $\delta = 10\%$.

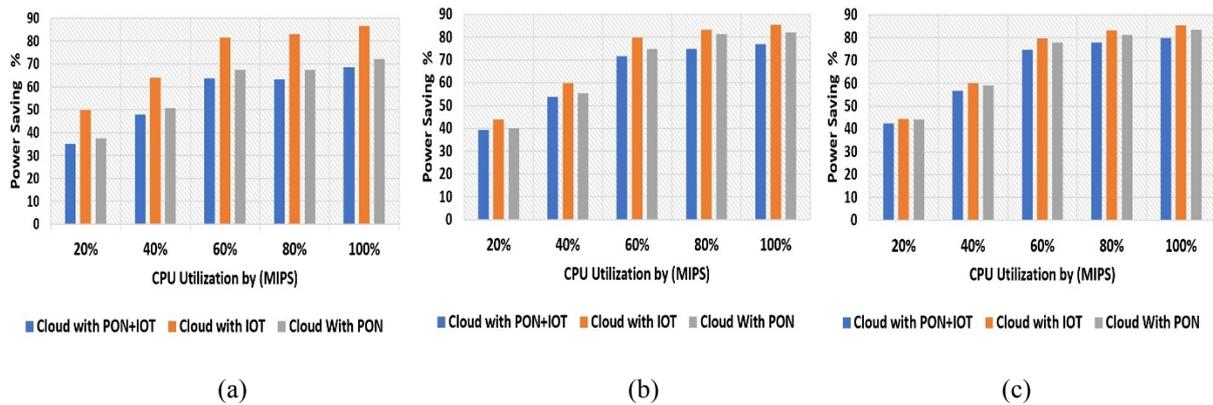

| (a) | (b) | (c) |

Figure 4. Total savings of the fog approach at different values of $\delta$: a) $\delta = 1\%$, b) $\delta = 5\%$, and c) $\delta = 10\%$.

## 4. CONCLUSIONS

This paper focused on developing a framework for energy efficient neural network (NN) embedding in IoT based smart home applications. The results use a MILP model which evaluates several processing placement solutions based on the concept of fog. A virtual topology of nodes and links represents the services to be embedded inline with the neural network process workflow and the service abstraction paradigm of SOA. The results show that the fog approach can produce power savings of up to 86% in all scenarios, given that the processing tasks are placed in IoT layer only. Generally, the fog approach despite the variation in the placement of tasks in different layers, produced substantial power savings compared to the centralized cloud solution.

ACKNOWLEDGMENTS
The authors would like to acknowledge funding from the Engineering and Physical Sciences Research Council (EPSRC), through INTERNET (EP/H040536/1), STAR (EP/K016873/1) and TOWS (EP/S016570/1) projects. The author would also like to thank the University of Tabuk for Education Development for funding his scholarship. All data is provided in the results section of this paper.